\documentclass[fleqn,10pt]{wlscirep}
\title{High-performance nanoscale topological energy transduction}

\author[1,2,*]{Timothy M. Philip}
\author[1,2]{Matthew J. Gilbert}
\affil[1]{University of Illinois at Urbana-Champaign, Department of Electrical and Computer Engineering, Urbana, IL 61801, USA}
\affil[2]{University of Illinois at Urbana-Champaign, Micro and Nanotechnology Laboratory, Urbana, IL 61801, USA}

\affil[*]{tphilip3@illinois.edu}

\usepackage{amsfonts}
\usepackage{bm}% bold math$
\usepackage{hyperref}
\hypersetup{
  linkcolor=blue,
  citecolor=blue,
  filecolor=blue,
  urlcolor=blue,
  colorlinks=true
}
\usepackage[export]{overpic}
\usepackage{xr}
\usepackage{epstopdf}
\renewcommand{\v}[1]{\ensuremath{\mathbf{#1}}} % vector
\newcommand{\vg}[1]{\ensuremath{\bm{#1}}}	% Greek vector
\newcommand{\abs}[1]{\left|{#1}\right|}

\begin{abstract}
	The realization of high-performance, small-footprint, on-chip inductors remains a challenge in radio-frequency and power microelectronics, where they perform vital energy transduction in filters and power converters. Modern planar inductors consist of metallic spirals that consume significant chip area, resulting in low inductance densities. We present a novel method for magnetic energy transduction that utilizes ferromagnetic islands (FIs) on the surface of a 3D time-reversal-invariant topological insulator (TI) to produce paradigmatically different inductors. Depending on the chemical potential, the FIs induce either an anomalous or quantum anomalous Hall effect in the topological surface states. These Hall effects direct current around the FIs, concentrating magnetic flux and producing a highly inductive device. Using a novel self-consistent simulation that couples AC non-equilibrium Green functions to fully electrodynamic solutions of Maxwell's equations, we demonstrate excellent inductance densities up to terahertz frequencies, thus harnessing the unique properties of topological materials for practical device applications.
\end{abstract}
\begin{document}

\flushbottom
\maketitle
\thispagestyle{empty}

\section*{Introduction}

On-chip inductors are integral, passive circuit components that convert or transduce electrical energy into magnetic energy for use in a variety of analog filter and voltage regulating circuits.\cite{Xu2011a,Balog2013,Morrow2011} Two key requirements for effective on-chip inductors are: a small footprint to allow for the integration of more active components within modern microelectronic chips and a high operating frequency as necessitated by the operating frequencies of modern and future device technologies. Due to the demands of the planar fabrication process, typical on-chip inductors consist of spiraled metallic wire traces, usually made from copper, that link the magnetic flux generated by the concentric coils to concentrate magnetic energy. Although the spiral geometry maximizes flux linkage for a two-dimensional system, limitations in fabricating highly-conducting, closely-spaced spiraled traces results in the consumption of a large chip area to create components with significant inductance.\cite{Wang2012a} Various solutions have been proffered to mitigate this issue from the incorporation of magnetic Ni$_{x}$Fe$_{1-x}$ yokes to enhance the magnetic field through the core\cite{Mathuna2005,Gardner2009,Wang2012a} to the substitution of graphene\cite{Sarkar2011,Zhou2013,Li2014} or carbon nanotubes\cite{Li2009} for the conducting material to increase the current flow within the coils. These solutions, however, are limited by their operating frequency, as is the case for magnetic yokes in copper inductors,\cite{Wang2012a} or by their fabrication reliability and low inductance density, for the carbon-based designs. Fundamentally, an inductor design based on new phenomena, geometries, and materials would enable on-chip inductors to achieve the size and inductance targets needed for nanoscale circuits of the future.~\cite{ITRS2015}

Recently discovered three-dimensional, time-reversal-invariant topological insulators (TIs) have drawn significant attention for possessing high mobility and for hosting novel physical phenomena.\cite{Butch2010,Hasan2010,Qi2011,Bernevig2013} A number of device applications using the unique properties of TIs ranging from transistors\cite{Cho2011,Zhu2013a,Li2014a} and interconnects\cite{Zhang2012,Philip2017} to more exotic applications such as spintronics\cite{Wu2011,Pesin2012,Duan2013} and quantum computation\cite{Fu2008} have been suggested, yet few have offered the performance or reliability necessary to be considered for integration into next-generation, post-CMOS electronic circuits. Like ordinary insulators, TIs have a bulk electronic band gap, but the nontrivial topology of their band structures results in gapless conducting two-dimensional Dirac fermions on their surface.\cite{Fu2007,Zhang2009a,Chen2009} Using the unconventional physics enabled by the Dirac surface states such as the anomalous Hall effect (AHE)\cite{Nagaosa2010a,Haldane2004} and the quantum anomalous Hall effect (QAHE),\cite{Haldane1988,Yu2010a} we present a pragmatically different geometry for magnetic energy transduction that does not rely on the conventional method of physically spiraling a conductor. We theoretically investigate the performance afforded by our topological inductor design by utilizing a novel hybrid quantum transport and electrodynamics simulation that captures the dynamic fields that enable flux linking. 

\section*{Results}

\subsection*{Device design and ideal operation}

\begin{figure}[t]
\begin{center}
  % {\includegraphics[height=0.14\textheight,clip,trim={0 1.2in 0 0}]{axis.jpg}}
  %   {\includegraphics[height=0.14\textheight]{Ek_Mz0_axis0_labelnu1_spin1_015deg.jpg}}
  %   \hspace{0.2em}
  %   \includegraphics[height=0.14\textheight]{Ek_Mz1_axis0_labelnu1_spin1_015deg.jpg}
  %   \hspace{0.em}
  %   \includegraphics[height=0.14\textheight]{Ek_Mz-1_axis0_labelnu1_spin1_015deg.jpg}
  %   % \vspace{0.5em}
  %   % {\centering\includegraphics[width=\columnwidth,clip,trim={0 4in 0.7in 0},page=5]{inductor_cartoon.eps}}
  %   %{\includegraphics[height=.12\textheight,clip,trim={.1in 4.6in 1.4in 0}]{QAHE_edge_state.pdf}}
  %   \hspace{0.2em}
  %   \raisebox{7mm}{\includegraphics[height=.12\textheight,clip,trim={.1in 12in 1.4in 0}]{QAHE_edge_state.jpg}}
  \includegraphics[height=.18\textheight]{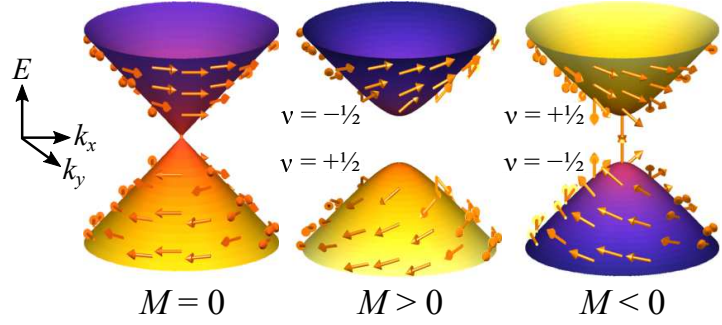}
\end{center}

\caption[Physics of topological insulator surface states]{\textbf{Physics of topological insulator surface states.}  The surface states of TIs host 2D Dirac surface states with linear dispersion when there is no magnetic Zeeman interaction ($M = 0$). Their characteristic spin-momentum locking is evident from the superimposed arrows indicating the spin. When magnetization is added with $M > 0$, a gap is generated creating two topologically nontrivial bands with Chern numbers $\nu = \pm 1/2$. When the sign of the magnetization is flipped with $M < 0$, the Chern numbers of the resulting bands also switch.  \label{fig:bandstructure}}
\end{figure}

The surface states of a TI are Dirac electrons characterized by the low-energy energy-momentum dispersion\cite{Zhang2009a}
% \begin{equation}
%   H_\text{surf} = k_x \sigma_y - k_y \sigma_x + M \sigma_z,
% \end{equation}{}
% where $k_i$ are the electron's momenta, $\sigma_i$ are the spin Pauli matrices, and $M$ is the magnetically-generated Zeeman energy. 
\begin{equation}
  E_\text{surf} = \sqrt{ \hbar^2 v_F^2 |\v k |^2 + M^2 },
\end{equation}
where $\hbar$ is the reduced Planck's constant, $v_F$ is the electron's Fermi velocity, $\v k$ is the electron's momentum, and $M$ is the magnetically-induced Zeeman energy. Figure~\ref{fig:bandstructure} illustrates the linear dispersion of the surface states in the absence of ferromagnetism when $M = 0$. The linear dispersion combined with the fact that spin, illustrated by the superimposed arrows in Fig.~\ref{fig:bandstructure}, is locked to momentum results in highly conductive surfaces with suppressed backscattering.\cite{Roushan2009} When a perpendicularly-oriented ferromagnet is placed in proximity to the surface resulting in $M \neq 0$, a gap opens in the dispersion that divides the surface states into topologically nontrivial 2D bands\cite{Liu2009,Chen2010,Xu2012} characterized by the Chern number $\nu$, as indicated in Fig.~\ref{fig:bandstructure}. When the magnetization orientation is away from (towards) the bulk, $M$ is positive (negative), resulting in the lower occupied band having a Chern number $\nu$ of +1/2 (-1/2). When an electric field is applied in a Chern insulating system while the chemical potential lies within the magnetic gap, charge is pumped perpendicular to the field by the QAHE with a quantized Hall conductivity $\sigma_{xy} = \nu_\text{occ.} e^2/h$, where $\nu_\text{occ.}$ is the sum of the Chern numbers of all occupied bands, $e$ is the electron charge, and $h$ is Planck's constant.\cite{Laughlin1981,Thouless1982a}

\begin{figure*}[t]
  \centering
  % \fbox{\includegraphics[width=0.8\textwidth,clip,trim={0.5in 0.7in 0.3in 0.8in},page=2]{inductor_cartoon.eps}}
  % {\includegraphics[width=\textwidth,clip,trim={0 3in 0.3in 0},page=3]{inductor_cartoon_wide.eps}}
  % \fbox{\includegraphics[width=\textwidth,clip,trim={0 3.2in 1.2in 0}]{inductor_cartoon_wide.pdf}}
  % {\includegraphics[width=\textwidth,clip,trim={0 13in 1.2in 0}]{inductor_cartoon_wide.jpg}}

  % {\includegraphics[width=0.31\textwidth]{{Fig2b_fdNEGF_IslandCurrent_mat3_Dope1_Mu0.10_freq1.00e+10}.jpg}} \hspace{.2em}
  % {\includegraphics[width=0.31\textwidth]{{Fig2c_fdNEGF_Bz_mat3_Dope1_Mu0.10_freq1.00e+10}.jpg} } \hspace{.2em}
  % {\includegraphics[width=0.31\textwidth]{Fig2d_InductanceFreqResp_3D_Poisson1.eps}}
  \includegraphics[width=\textwidth]{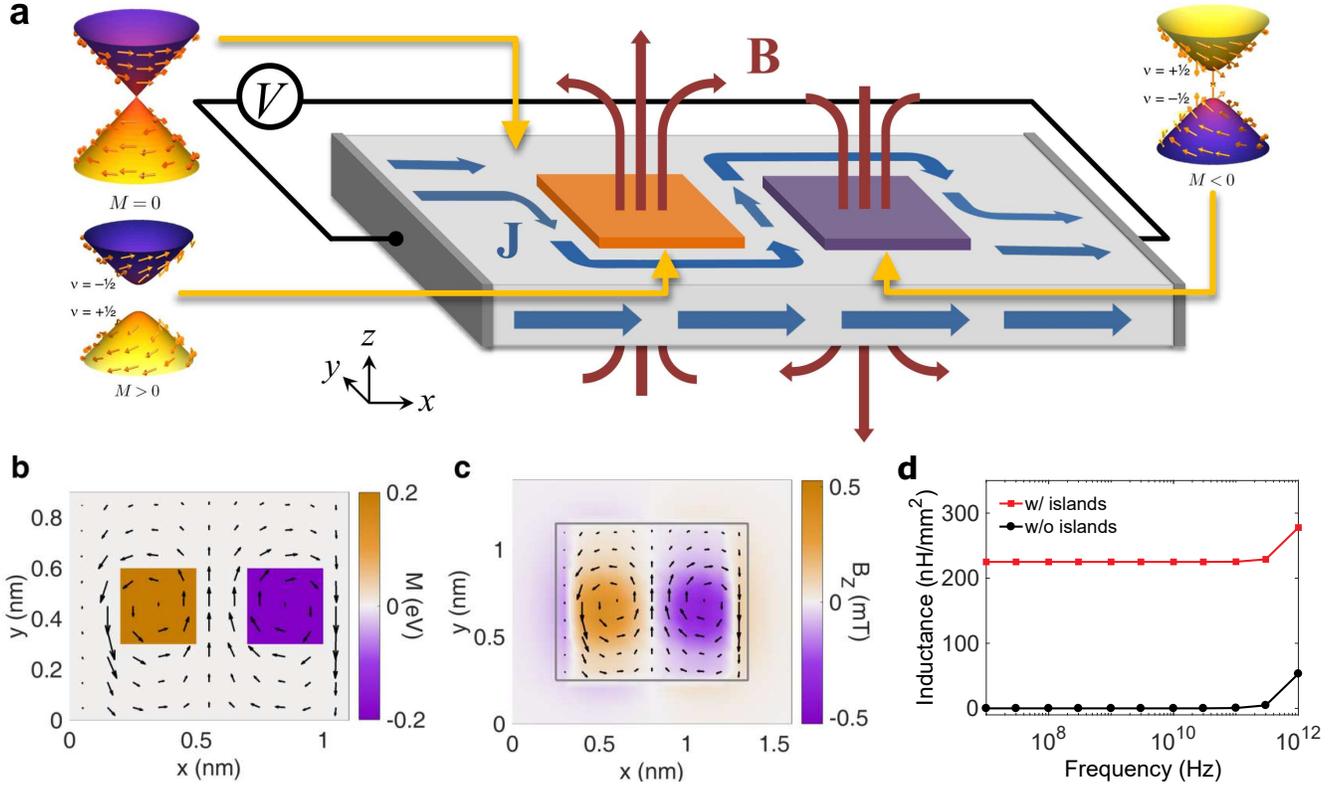}
  \caption[Schematic and ideal operation of the topological inductor]{\textbf{Schematic and ideal operation of the topological inductor.} \textbf{(a)} A schematic of a two-island topological inductor that utilizes the surface states of a time-reversal-invariant TI. By alternating the magnetization of each subsequent island, indicated by the orange and purple squares, under bias, the surface current density wraps around the FIs because of the QAHE induced by the ferromagnetism. The altered motion of the current around the islands concentrates magnetic energy through the islands, resulting in enhanced inductance.  \textbf{(b)}  The self-consistent AC NEGF simulation of the the current density in the top surface of a topological inductor under an AC bias of 10 mV reveals the circulating currents generated by the QAHE. The FIs are indicated by the two colored squares where the Zeeman field $M$ is nonzero, and the current density by the black arrows. \textbf{(c)} The resulting $\hat z$-directed magnetic flux density $B_z$ at an AC bias of 10 mV, where the gray outline indicates the position of the inductor. The electromagnetic domain is larger than the NEGF domain to capture any fringing and radiating fields. \textbf{(d)} The frequency response of the topological inductor demonstrates high inductance density up to terahertz frequencies.\label{fig:design}}
\end{figure*}

Figure~\ref{fig:design}a illustrates how this unique Hall response can be utilized to make a highly-efficient topological inductor. The design involves a TI substrate where the chemical potential is within the bulk band gap resulting in transport being carried solely through the surface states. Ferromagnetic islands (FIs), indicated as orange and purple squares corresponding to $+\hat{\hat{z}}$ and $-\v{\hat{z}}$ oriented magnetizations, respectively, are placed on the surface of the TI to selectively create magnetic band gaps in the surface state dispersion. For ideal operation, the chemical potential is placed within the magnetic band gap such that the ferromagnetically-doped regions are insulating. When the surface current density $\v{J}$, generated by a bias $V$ applied in the $\v{\hat x}$ direction, encounters the first island with $M > 0$ and $\nu_\text{occ.}= +1$, it is guided counter-clockwise around the island by the QAHE. After traversing around the first island, the surface current is then directed clockwise around the second island with $M < 0$ and $\nu_\text{occ.} = -1$ by the opposite flowing QAHE. By directing the current density around the islands, the current-generated magnetic flux density $\v B$ is concentrated through the FIs resulting in the storage of magnetic energy. The magnetic fields generated by circulating currents around an FI, in addition to that created by the currents encircling nearby FIs, create flux linkages that amplify the magnetic energy within the system and result in a highly inductive device.

%%%%%%%%%%%%%%%%%%%%%%%%%%%%%%%%%%%%%%%%%%%%%%%%%%%%%%%%%%%%%%%%%%
%%%%    Figure 1: Chern Insulator Bandstructure
%%%%%%%%%%%%%%%%%%%%%%%%%%%%%%%%%%%%%%%%%%%%%%%%%%%%%%%%%%%%%%%%%%
%%%%%%%%%%%%%%%%%%%%%%%%%%%%%%%%%%%%%%%%%%%%%%%%%%%%%%%%%%%%%%%%%%
%%%%%%%%%%%%%%%%%%%%%%%%%%%%%%%%%%%%%%%%%%%%%%%%%%%%%%%%%%%%%%%%%%

We theoretically evaluate the basic implementation and efficacy of the topological inductor by simulating the device using a novel method that couples AC quantum transport self-consistently with the full solution of Maxwell's equations for electrodynamics in three-dimensions. A fully quantum treatment of transport is necessary to capture the topological QAHE that is integral to the device operation. We use the AC non-equilibrium Green function (NEGF) technique that computes the first-order response of a device to an AC driving voltage (See Methods section).\cite{Wei2009,Kienle2010,Zhang2013} The computed currents and charge density from AC NEGF are then input into a fully dynamic finite-difference frequency-domain (FDFD) electromagnetics simulation\cite{Luebbers1990} to accurately evaluate the inductance resulting from the dynamic magnetic flux generation (See Methods section). The output electrodynamic potentials are then input back into the transport equations resulting in a iterative cycle that is terminated once the change in dynamic potentials between successive iterations is less than 1 $\mu$V, which we define as our criterion to reach self-consistency. It is important to note that a simple self-consistent solution of Poisson's equation severely underestimates the inductance even at low frequencies, thus demonstrating the necessity for the full dynamic electromagnetic calculation to capture the flux linking by the circling currents (See Supplementary Note~1 and Supplementary Fig.~1). While it is understood that current through the surface states of TIs can generate a strong spin-transfer torque that can alter magnetization direction,\cite{Fan2014,Mellnik2014,Fischer2016} we assume the FIs have high coercivity and, thus, negligible magnetization dynamics.

%%%%%%%%%%%%%%%%%%%%%%%%%%%%%%%%%%%%%%%%%%%%%%%%%%%%%%%%%%%%%%%%%%
%%%%    Figure 2: Inductor Design
%%%%%%%%%%%%%%%%%%%%%%%%%%%%%%%%%%%%%%%%%%%%%%%%%%%%%%%%%%%%%%%%%%
%%%%%%%%%%%%%%%%%%%%%%%%%%%%%%%%%%%%%%%%%%%%%%%%%%%%%%%%%%%%%%%%%%
%%%%%%%%%%%%%%%%%%%%%%%%%%%%%%%%%%%%%%%%%%%%%%%%%%%%%%%%%%%%%%%%%%

Using this coupled AC NEGF-FDFD technique, we simulate a ($12a_0$, $10a_0$, $5a_0$) device, where $a_0 = 1$ \AA, with a model Hamiltonian that reproduces the same symmetries of a 3D TI and has a bulk band gap of 1 eV (See Methods section).  Square FIs are placed on the top surface with side length of 0.3 nm, separation of 0.2 nm, and  $M = \pm 0.2$ eV.  The chemical potential is set to 0.1 eV to be within the magnetic gap such that the inductor operates within the QAHE regime. The temperature is set to 300 K, but results are largely insensitive to the specific choice as long all relevant energy scales are well above the thermal energy. After self-consistency is attained, the inductance $L$ is calculated as $L = 2 E_B/I^2$, where $I$ is the current through the device and the stored magnetic energy $E_B$ is calculated as $E_B = \int dV \frac{1}{\mu_0}\abs{\v B}^2$, where $\mu_0$ is the magnetic permeability of the material and $\v B$ is the magnetic flux density. 

Figure~\ref{fig:design}b displays the AC current density profile of the top surface of the device at a frequency of 10 GHz and AC voltage of 10 mV. Since the AC observables are averaged over the period of the driving frequency, the resultant current density appears to completely encircle the islands due to the addition of forward and backward current flow. This current circulation due to the QAHE generates high magnetic fields over the islands as shown in Fig.~\ref{fig:design}c. In Fig.~\ref{fig:design}d, we repeat the simulation of the topological inductor over a frequency range from 10 MHz to 1 THz. Without any specific optimization of the device geometry, we achieve an inductance density of 225 nH/mm$^2$, an order of magnitude greater than the 23.2 nH/mm$^2$ attained by CNT inductors and comparable to the 1000 nH/mm$^2$ of high-density copper spiral inductors. The topological inductor sustains this performance over the entire frequency range simulated, which is well above the low cut-off frequencies, ranging from 0.2 GHz to 150 GHz, of other current and proposed designs. When we simulate a bare TI without the FIs, we find that the surface states naturally have an inductance density less that is than one pH/mm$^2$, which demonstrates the dramatic increase the FIs can have on energy transduction. At high frequencies near 1 THz, we observe an increase in the inductance both with and without islands. At such high frequencies, spurious charge accumulation due to the AC NEGF contact approximation utilized results in an artificial increase in the inductance.\cite{Zhang2013} In principle, however, the only limitation on operation frequency is the size of the island compared to the wavelength of the driving voltage. Once the island side length exceeds half a wavelength, the rapidly oscillating electric field does not produce uniform circulating currents around the islands, resulting an unreliable current density and magnetic field profile.

\subsection*{Non-idealities}

%%%%%%%%%%%%%%%%%%%%%%%%%%%%%%%%%%%%%%%%%%%%%%%%%%%%%%%%%%%%%%%%%%
%%%%    Figure 3: fdNEGF transport
%%%%%%%%%%%%%%%%%%%%%%%%%%%%%%%%%%%%%%%%%%%%%%%%%%%%%%%%%%%%%%%%%%
\begin{figure*}[t]
% \begin{center}
 %  {\includegraphics[width=0.25\textwidth]{Fig3a_SpaceCurrentResp.eps} } \hspace{-0.5em}
 %  % \raisebox{2mm}{\includegraphics[width=0.25\textwidth]{Fig3b_DCInductanceMuResp_3D_Poisson1.eps} } \hspace{-1.5em}
 %   {\begin{overpic}[width=0.25\textwidth]{Fig3b_InductanceMuResp_3D_Poisson1.eps}
 %      \put(20,16){{\includegraphics[width=0.067\textwidth,clip,trim={0 5in 11.25in 0 in}]{TI_band_cartoon.eps}}} 
 %      \end{overpic}}\hspace{-1.1em}
 %  % \begin{overpic}[width=0.25\textwidth,clip,trim={1.4in 3.3in 1.5in 3.0in}]{Fig3c_BzMuResp_3D.eps}
 %  \begin{overpic}[width=0.25\textwidth]{Fig3c_BzMuResp_3D.eps}
 %      \put(50,49){{\includegraphics[width=0.1\textwidth,clip,trim={0 5.48in 5.8in 0 in}]{y0.eps}}} 
 %      \end{overpic}  \hspace{-1em}
 %  % {\includegraphics[width=0.25\textwidth,clip,trim={1.4in 3.3in 1.6in 3.in}]{Fig3d_JyMuResp_3D.eps}} \hspace{-1.5em}
 %  {\includegraphics[width=0.25\textwidth]{Fig3d_JyMuResp_3D.eps}} \hspace{-1.5em}
  
 % \vspace{0.5em}

 %  \raisebox{6.25mm}{\includegraphics[width=0.275\textwidth]{Fig3e_InductanceDisorderResp.eps} }  \hspace{-1.6em}
 %  \raisebox{6.0mm}{\includegraphics[width=0.275\textwidth]{{Fig3f_fdNEGF_DisorderSites_10x10x5_mat3_muD1.00e-01_doping1_Disorder0-3.00e-01}.jpg} } \hspace{-1.6em}  
 %  \raisebox{4.25mm}{\includegraphics[width=0.18\textwidth,clip,trim={0.2in 5in 20in 0in}]{skew_scattering.jpg} } %\hspace{0em}
 %  {\includegraphics[width=0.25\textwidth]{{Fig3h_MagIslandDopant_Mz-0.20_D+1.74_Sz}.jpg} } 
 \includegraphics[width=\textwidth]{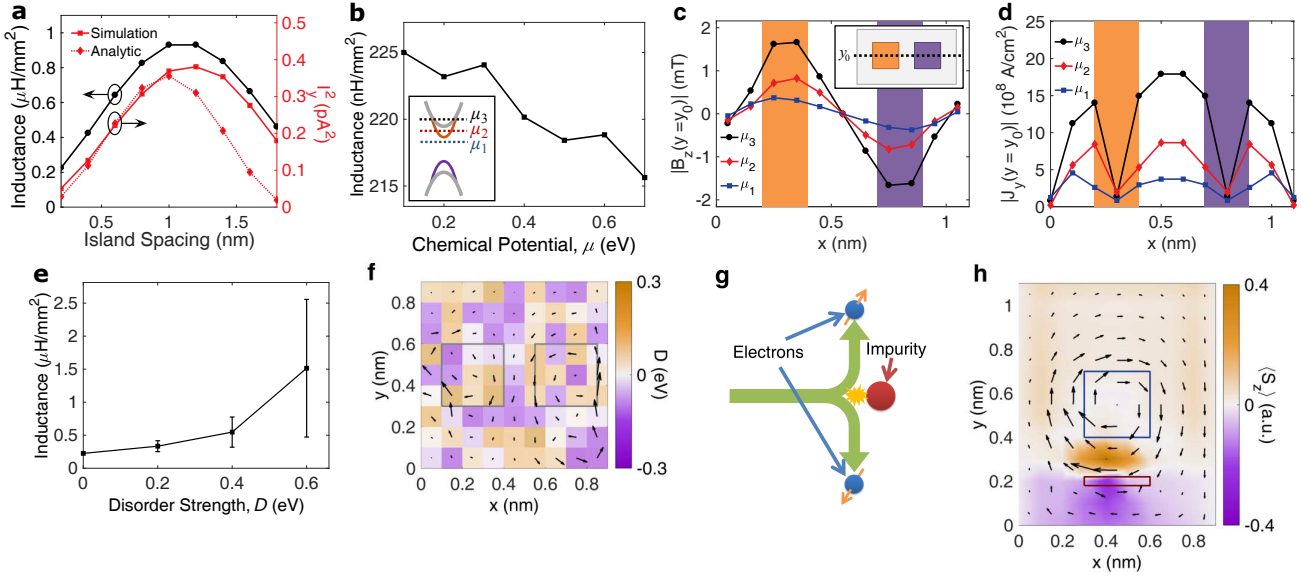}

% \end{center}
  \caption[Device performance with respect to island spacing, chemical potential, and impurity disorder]{\textbf{Device performance with respect to island spacing, chemical potential, and impurity disorder.} \textbf{(a)} The inductance is maximized when a half-period of the input voltage signal is captured between the FIs and is proportional to the square of the current between the FIs, $I_y^2$, which is calculated numerically and approximated analytically. \textbf{(b)} Inductance at a frequency of 10 GHz as a function of chemical potential. The locations of $\mu_1 = 0.1$ eV, $\mu_2 = 0.3$ eV, and $\mu_3 = 0.6$ eV, which lie within the magnetic gap, inside only the surface bands, and inside both the surface and bulk bands, respectively, are illustrated on the band structure schematic in the inset. \textbf{(c)} The magnetic field $B_z$ as a function of $x$ at $y = y_0$ (see inset device schematic) increases with chemical potential. The shaded regions correspond to the location of FIs. \textbf{(d)} The current density $J_y$ at $y = y_0$ reveals that the the stronger magnetic field is due to the increase in circulating current around the islands. \textbf{(e)} Inductance of the design under the influence of random disorder impurity potentials. \textbf{(f)} The current density on the surface of the device is illustrated at disorder strength of $D=0.6$ eV, where the square outlines mark the location of the FIs. As disorder is increased, the current density becomes dominated by circulating currents caused by skew scattering off impurities. \textbf{(g)} A schematic of skew scattering, whereby electrons with different spins scatter to different directions due to an asymmetry in the scattering amplitude of a given spin, generated by broken time-reversal symmetry. \textbf{(h)} Skew scattering is easily recognized in a simulation of a single magnetic island, indicated by the blue square with $M = -0.2$ eV, and a nearby row of impurities with strength 1.7 eV, indicated by the red rectangle. The characteristic rotation of current around the FIs is disrupted  by skew scattering,  resulting in the accumulation of spin on each side of the impurities. \label{fig:nonidealities}} 
\end{figure*}

%%%%%%%%%%%%%%%%%%%%%%%%%%%%%%%%%%%%%%%%%%%%%%%%%%%%%%%%%%%%%%%%%%
%%%%%%%%%%%%%%%%%%%%%%%%%%%%%%%%%%%%%%%%%%%%%%%%%%%%%%%%%%%%%%%%%%

The performance of the topological inductor is intimately tied to the flux linking between adjacent current loops and therefore is a function of the spacing between successive islands. Thus, we simulate a larger ($32a_0$, $10a_0$, $5a_0$) structure with 0.3 nm side length islands, a frequency of 10 GHz, and island spacing ranging from 0.2 nm to 1.7 nm to understand the effect of island spacing on the observed inductance. In Fig.~\ref{fig:nonidealities}a, we plot the inductance in addition to the numerically calculated and analytically approximated square of the current between the FIs, $I_y^2$, (See Supplementary Note~2 and Supplementary Fig.~2). The magnetic field generated by the circulating current is proportional to current density by Amp\'ere's law, and inductance is proportional to the square of the magnetic field. Therefore, it follows that the peak inductance can be found by maximizing the amount of current that circulates around and between the islands, which is proportional to $I_y$. The analytic approximation for the current between the islands, approximated as semi-infinite magnetic regions, reveals that the current between the islands varies sinusoidally with island spacing. The inductance peaks at the width that captures a half period of the sinusoidal current profile, which corresponds a spacing of 1.2 nm in our system. After 1 nm, the islands can no longer be approximated as semi-infinite, and thus the analytic calculations begins to deviate from the numerical.

The chemical potential, $\mu$, in TIs cannot always be accurately placed within the magnetic band gap. For example, in the TI Bi$_2$Se$_3$, the proliferation of selenium vacancies in the growth process results in a highly electron-doped material in which the chemical potential crosses the bulk bands.\cite{Sun2010} Controlling the chemical potential position in these materials with conventional methods such as electrostatic gating has proved to be challenging\cite{Xia2009} and thus it is imperative to understand how its position effects the resultant inductance. Figure~\ref{fig:nonidealities}b shows that the inductance in fact decreases with increasing $\mu$ at a frequency of 10 GHz. To better understand the device response to the chemical potential placement, we consider three values depicted in the inset: $\mu_1 = 0.1$ eV, which lies within the magnetic band gap, $\mu_2 = 0.3$ eV, which crosses the surface band, and $\mu_3 = 0.6$ eV, which crosses both the bulk and surface bands. Figure~\ref{fig:nonidealities}c shows a cross section along the center of the device, $y = y_0$ in the inset, of the $\v{\hat z}$ magnetic flux density, $B_z$, with the location of the FIs illustrated by orange and purple shaded regions. We see that the magnetic flux density grows with increasing chemical potentials $\mu_2$ and $\mu_3$. Figure~\ref{fig:nonidealities}c similarly displays a cross cut of the the $\v{\hat y}$ current density, $J_y$, and reveals that the the increased magnetic flux density is due to the increase in current circulating around the islands. When the chemical potential crosses the surface bands at $\mu_2$, the QAHE is replaced by the AHE, whereby a bulk, non-quantized transverse current is generated by a longitudinal electric field.\cite{Nagaosa2010a} The AHE combined with increased electron density at the higher chemical potentials results in the larger current densities observed. The enhancement of the magnetic flux density and circulating current density, however, is offset by an increase in terminal current, resulting in the net decrease of inductance seen in Fig.~\ref{fig:nonidealities}b. Raising $\mu$ further to $\mu_3$, where bulk states contribute to transport, results in a further reduction in inductance as the terminal current is again increased. Despite this non-ideal current flow resulting from the AHE and additional conduction through the bulk states, we observe only a 4\% reduction in inductance at $\mu = 0.7$ eV, indicating that the performance of the inductor is largely independent of the specific location of the chemical potential.

Conduction through the surface states of TIs is known to be robust to the presence of non-magnetic disorder,\cite{Roushan2009,Alpichshev2010} but time-reversal-breaking ferromagnetism can destroy this topological protection.\cite{Zhang2012a} Since the operation of the topological inductor is reliant on the presence of FIs that break time-reversal symmetry, the surface states may not be as resilient to disorder as a pristine TI sample. To characterize the influence disorder has on the topological inductor performance, we calculate the inductance with the original dimensions of ($12a_0$, $10a_0$, $5a_0$) at a frequency of 10 GHz with $\mu = 0.1$ eV and include the presence of on-site impurity potentials throughout the device domain with energies ranging between $-D/2$ and $D/2$, where $D$ is the disorder strength. Figure~\ref{fig:nonidealities}e shows the simulated inductance averaged over three real-space disorder potential configurations as a function of the disorder strength. We find that the inductance rises with disorder strength, leading to an inductance density of 1.5 $\mu$H/mm$^2$ at $D=0.6$ eV. We also observe that the variance in inductance increases with disorder strength, indicating that the physical layout of dopants plays a considerable role in the resulting energy transduction. Figure~\ref{fig:nonidealities}f shows the current density profile of the top layer of the device overlaid on a specific disorder potential profile distribution with $D = 0.6$ eV. Rather than observing the currents circulating around the FIs, we see a much more erratic current distribution. The disorder strengths studied here are lower than that which would be necessary for a disorder-induced phase transition.\cite{Kobayashi2013} Thus, the disturbances observed are related to scattering in the now vulnerable surface states.  The origin of such unpredictable electron motion can be traced back to the onset of skew scattering off the impurity potentials. Skew scattering, illustrated schematically in Figure~\ref{fig:nonidealities}g, is a spin-selective scattering mechanism that, although unrelated to topological character, is unique to systems possessing strong spin-orbit coupling and broken time-reversal symmetry, as considered here.\cite{Smit1958,Vignale2010}  In skew scattering, the time-reversal-breaking magnetic field or magnetization generates an asymmetry in the scattering transition probability based on the spin of the electron.\cite{Sinitsyn2007,Nagaosa2010a,Xiao2010} Therefore, an spin-up electron deflects off an impurity in the opposite direction of a spin-down electron. Since the surface states of the TI are spin-momentum locked, right-moving electrons have opposite spin of left-moving electrons and therefore scatter in opposite directions. Figure~\ref{fig:nonidealities}h shows this clearly in a simulation of a single FI with $M = -0.2$ eV marked with the blue square and a row of impurity potentials with barrier height 1.7 eV marked by the red rectangle. The physical origin of the asymmetric scattering is easily understood in this example: a left-moving electron encountering the impurities will more likely deflect above them into the circulating QAHE current around the FI. Similarly, it is energetically unfavorable for right-moving electrons to scatter into the opposite-moving QAHE current, so they scatter below the impurities. Because left-moving and right-moving electrons take opposite paths around the dopants, the AC current density distribution in Figure~\ref{fig:nonidealities}h appears to encircle the impurities. The expectation value of $\v{\hat z}$ spin, $\langle S_z \rangle$, accumulates with opposite sign on each side of the impurities, which demonstrates that the underlying mechanism for this disturbance to the current profile is indeed skew scattering. The inadvertent current circulation around impurities due to skew scattering causes localized magnetic flux ``pockets'' away from the magnetically defined regions, resulting in a net increase in the inductance of the device. However, because these impurities are randomly placed these calculations show that although disorder does not degrade performance in the topological inductor, it does make the inductance more difficult to predict due to the loss of control of the surface current density distribution.

%%%%%%%%%%%%%%%%%%%%%%%%%%%%%%%%%%%%%%%%%%%%%%%%%%%%%%%%%%%%%%%%%
%%%    Table 1: Comparison to other technologies
%%%%%%%%%%%%%%%%%%%%%%%%%%%%%%%%%%%%%%%%%%%%%%%%%%%%%%%%%%%%%%%%%
\begin{table}[t]
   \centering
     \begin{tabular}{| l | c | c |}
      \hline
       							  		& Cut-off 		  	& Inductance \\
      Inductor                    		& Frequency (GHz) 	&  (nH/mm$^2$) \\
      \hline
      LF Copper\cite{Gardner2009} 		& 0.2             	& 1700        \\ 
      RF Copper\cite{Xu2011a}     		& 6            		& 282         \\ 
      CNT\cite{Li2009}            		& 150             	& 23.2        \\ 
      Graphene\cite{Sarkar2011}   		& 150             	& 636         \\ 
      Topological Inductor        		& 1000            	& 930      \\ \hline
    \end{tabular}  
    \caption[Comparison of modern inductor performance]{\textbf{Comparison of modern inductor performance.}{ Low-frequency (LF) copper-based inductors provide a large inductance due to their low resistance, but this performance is limited below one GHz due to the skin effect that constricts current flow. Higher frequency radio-frequency (RF) copper inductors can offer higher cut-off frequencies at the cost of a significantly lower inductance density. Carbon-based CNT and graphene designs offer moderate and high inductance, respectively, but their operation frequency is again limited by the anomalous skin effect that greatly increases resistance above 150 GHz. As the topological inductor utilizes surface current flow, skin effects have negligible impact on performance and thus offers high inductance into terahertz frequencies.}\label{tab:comparison}}
\end{table}
%%%%%%%%%%%%%%%%%%%%%%%%%%%%%%%%%%%%%%%%%%%%%%%%%%%%%%%%%%%%%%%%%
%%%%%%%%%%%%%%%%%%%%%%%%%%%%%%%%%%%%%%%%%%%%%%%%%%%%%%%%%%%%%%%%%

\section*{Discussion}

In order to benchmark the topological inductor for use as an on-chip inductor, we compare its performance to current and proposed inductor designs in Table~\ref{tab:comparison}. Although the physical dimensions of our simulated device are small, by comparing inductance per unit area, we obtain metrics that are independent of the device geometry, thereby allowing us to compare different technologies on equal footing. The low resistance of copper combined with recent advances in depositing magnetic yokes to enhance magnetic flux linking gives copper inductors superior low-frequency performance exceeding 1700 nH/mm$^2$.\cite{Gardner2009} This high inductance density, however, is limited to below one GHz. At high frequencies, the skin affect constricts current to the surface of the copper wire, dramatically increasing resistance and rapidly decreasing the inductance below 40 nH/mm$^2$.\cite{Wang2012a,Gardner2009} Radio-frequency copper inductors can be offer reliable performance up to 6 GHz, but their inductance density is greatly reduced to 282 nH/mm$^2$ due to a combination of skin effect resistance increases and ferromagnetic resonance permeability degradation.\cite{Xu2011a} High-mobility carbon-based conductor materials have been proffered as alternatives to copper-based design and have dramatically increased cut-off frequencies of up to 150 GHz. To create carbon-nanotube (CNT) spiral inductors, a metallic contact must be placed at each turn of the design, resulting in a high series contact resistance that severely restricts inductance densities below 23.2 nH/mm$^2$. Since graphene-based design can be lithographically patterned, their inductance is not limited by a series contact resistance, like CNT designs, and thus can reach inductance densities in excess of 600 nH/mm$^2$. The anomalous skin affect, an analog of the normal skin affect relevant in materials with mean free paths longer than the skin depth, however, limits the conductance of graphene inductors beyond 150 GHz.\cite{Sarkar2011,Li2014} The novel, simple geometry of the topological inductor allows it to achieve an inductance density of 930 nH/mm$^2$, approaching inductance densities of state-of-the-art magnetic-core copper inductors at operating frequencies well above those of competing technologies. This broad spectrum performance is afforded by the fact that its operation is based on surface conduction. Therefore, any high frequency surface confinement effects do not change conduction properties and the inductance is unaltered. Furthermore, as we are not concerned with motion of the ferromagnetic domains, we are not constrained by the known high-frequency limitations associated with ferromagnetic resonances.\cite{Gardner2009}

While our proof-of-concept inductor design demonstrates high performance, greater inductance may be achieved by adding more islands in series, thereby would increase flux linkages between islands. Additionally, further flux linking can be generated by adding islands to the bottom surface and side walls. The ability to optimize island size, spacing, and arrangement makes this system a versatile promising inductor design. As the operation of the inductor only requires the presence of a QAHE or AHE, the design is not limited only to the surface of TIs and can be realized in a variety of material systems including but not limited to Weyl semimetals\cite{Yang2011,Burkov2014,Zyuzin2016a}, 2D transition-metal dichalcogenides,\cite{Parkin1980,Cai2013,Mak2014} and dilute magnetic semiconductor systems.\cite{Bruno2004,Oveshnikov2015} 

Our study illustrates that the unique properties of TIs provide a platform for novel information processing device architectures. By placing ferromagnetic islands with alternating magnetization on the surface of a TI, we utilize the QAHE or AHE to deform the current density around the islands, concentrating magnetic flux within current loops. When simulated with a hybrid AC quantum transport and frequency-domain electromagnetics simulation, we find that the topological inductor offers high performance over a broad frequency range, making it an exceptional candidate for use in nanoscale wireless communication and power electronic applications.

\section*{Methods}

\subsection*{Model Hamiltonian}
The systems are modeled by a tight-binding Hamiltonian with nearest-neighbor hopping, which is given by
\begin{equation}
  \mathcal H(\v r) = \sum_{\v{r}} \left[ \psi_{\v{r}}^\dag H_0\psi_{\v{r}} + \sum_{\vg\delta}\left(\psi_{\v{r}}^\dag H_{\vg\delta}\psi_{\v{r} + \vg\delta} + \text{H.c.}\right) \right], \label{eq:TBHam}
\end{equation}
where $\psi_{\v{r}}$ is the electron annihilation operator, $\vg\delta = (\pm a \,\v{\hat x}, \pm a \,\v{\hat y}, \pm a \,\v{\hat z})$ are the distances between nearest neighbor atoms on a cubic lattice with lattice constant $a= 1$ \AA. In Eq.~\ref{eq:TBHam}, $H_0$ is the on-site term, and $H_{\v\delta}$ is the nearest-neighbor hopping term. The three-dimensional TI Hamiltonian requires a basis of two orbital and two spins resulting in the on-site term:~\cite{Liu2010a}
\begin{equation}
  H_0 = \mathbb M \Gamma^0 + M(\v r)\Gamma_M - eV(\v r) I_4,
\end{equation}
where $\Gamma^0 = \tau^z \otimes I_2$, $\Gamma_M = I_2 \otimes \sigma^z$, $\tau^i$ are the orbital Pauli matrices, $\sigma^i$ are the spin Pauli matrices, $I_N$ are the $N\times N$ identity matrices, and $\mathbb M = m - 3b/a^2$. Here, $m$ and $b$ are parameters that can be tuned to fit characteristics of a time-reversal-invariant 3D topological insulator. The spatially-varying Zeeman field generated by a surface-perpendicular ferromagnet is added through $M(\v r)$, and the scalar electromagnetic potential profile is incorporated through $V(\v r)$. The hopping term for this model is given by
\begin{equation}
  H_{\v\delta} =  \frac{b\Gamma^0 + i\gamma\, \vg\delta \cdot \vg\Gamma}{2a^2} \exp\left(\frac{ie}{\hbar} \int_{\v r}^{\v r + \v\delta} \v A(\v r) \cdot d\v\ell\right).
\end{equation}
Here, $\Gamma^{i} = \tau^x \otimes \sigma^i \, (i \in \{x, y, z\})$, $\vg\Gamma =  (\Gamma^x\;\v{\hat{x}},\Gamma^y\;\v{\hat{y}},\Gamma^z\;\v{\hat{z}})$, $e$ is the electron charge, $\hbar$ is the reduced Planck's constant, and $\gamma$ is an additional tunable parameter. The vector potential $\v A(\v r)$ enters through the Peierl's phase in this hopping term.\cite{Graf1995}

This model Hamiltonian reproduces the low energy physics of a TI including the anomalous Hall effect (AHE) and the quantum anomalous Hall effect (QAHE) by preserving the same symmetries of a time-reversal-invariant TI and obeying the proper Clifford algebra.\cite{Liu2010a} To understand the qualitative transport features of a TI, we set $m = {1.5}$~eV, $b={1}$~eV$\cdot$\AA$^2$, and $\gamma = {1}$~eV$\cdot${\AA} to create a bulk band gap of 1 eV that highly localizes the surface states such that they do not hybridize even at nanometer dimensions.

\subsection*{AC NEGF} The DC NEGF formalism has found great success in modeling fully quantum mechanical electron transport in nanoscale devices,\cite{Lake1997,Anantram2008a} and recent theoretical advances have extended the method to small-signal AC biases.\cite{Wei2009,Kienle2010,Zhang2013} The retarded Green function, that is, the impulse response of the system Hamiltonian, at energy $E$ can be expressed as
\begin{equation}
  G^r(E) = G^r_0(E) + G^r_\omega(E), \label{eq:Gdef}
\end{equation}
where $G^r_0(E)$ is the DC retarded Green function and $G^r_\omega(E)$ is first-order response due to an AC perturbation. The DC component is calculated via the standard NEGF formalism as \cite{Lake1997,Anantram2008a}
\begin{equation}
  G^r_0(E) = \left[ E - i\eta - \mathcal H - U - \Sigma_0^r(E) \right]^{-1},
\end{equation}
where $U$ is the static potential energy profile, $\eta$ is an infinitesimal positive number that pushes the poles of the Green function into the complex plane, allowing for integration along the real energy axis,\cite{Anantram2008a} and $\Sigma_0^r(E)$ is the contact self-energy that integrates out the influence of the semi-infinite leads. We assume the wide bandwidth limit (WBL) where the contacts have a much larger bandwidth than the device with a constant density of states as a function of energy. This assumption results in a retarded self-energy of the form $\Sigma_0^r(E) = i \Gamma$, where $\Gamma$ is the energy level broadening introduced by the leads.  As the contacts are typically  much larger than the device region, the WBL is a valid assumption as the number of available states in the lead should not vary greatly over biases and frequencies much less than the bandwidth of the metallic contact.

Since the AC bias is introduced perturbatively, the small-signal retarded AC Green function $G^r_\omega(E)$ at frequency $\omega$ is expressed as a product of DC Green functions at energies $E$ and $E+\hbar\omega$:\cite{Wei2009} 
\begin{equation}
  G^r_\omega(E) = G^r_0(E+\hbar\omega)\left[ -eV(\omega) + \Sigma^r_\omega(E+\hbar\omega, E)  \right] G^r_0(E).
\end{equation}
Here  $V(\omega)$ is the AC potential profile and $\Sigma^r_\omega$ is the AC contact self-energy. Just as the AC Green function is the small-signal perturbation to the DC Green function, the total contact self-energy can be expressed as 
\begin{equation}
    \Sigma^\gamma(E) = \Sigma^\gamma_0(E) + \Sigma^\gamma_\omega(E) \quad (\gamma = r, <), \label{eq:Sigmadef}
\end{equation}
where $\Sigma^\gamma_\omega(E)$ is the AC self-energy due to a perturbative bias of the form $V(t) = V_{AC} \cos \omega t$, where $V_{AC}$ is the amplitude of the AC driving voltage. The AC contact self-energy is similarly a function of the DC contact self-energies and is calculated as
\begin{equation}
  \Sigma^\gamma_\omega(E) = \frac{eV_{AC}}{\hbar\omega} [\Sigma_0^\gamma(E) - \Sigma_0^\gamma(E+\hbar\omega)] \quad (\gamma = r, <), \label{eq:sigma}
\end{equation}
where $e$ is the electron charge. In the WBL, the AC retarded self-energy greatly simplifies to $\Sigma^r_\omega= 0$.  Although the WBL provides an accurate description of large reservoir contacts for small energy scales, it neglects the nontrivial energy dependence of the contact self-energy at high frequencies where $\hbar\omega$ is no longer small and $[\Sigma_0^\gamma(E) - \Sigma_0^\gamma(E+\hbar\omega) ] \not\approx 0$, which can result in unphysical charge accumulation.\cite{Zhang2013} However, for small frequencies, where $\hbar\omega \ll 1$ and $[\Sigma_0^\gamma(E) - \Sigma_0^\gamma(E+\hbar\omega) ] \approx 0$, the WBL can be safely applied to model the AC self-energy of metallic leads.

In order to account for the application of a bias, the retarded Green function must be convolved with the lesser self-energy $\Sigma^<(E)$, which accounts for the occupancy of the leads, using the Keldysh equation $G^<(E) = G^r(E) \Sigma^<(E) G^r(E)^\dag$.\cite{Wei2009} After applying the definitions in Eqs.~\ref{eq:Gdef} and~\ref{eq:Sigmadef} and taking only the terms that are first-order in the perturbation, we obtain the expression for the AC lesser Green function:
\begin{equation}
        G_\omega^<(E) = G_0^r(E+\hbar\omega)\Sigma_0^<(E+\hbar\omega) G_\omega^r(E)^\dag 
                        + G_0^r(E+\hbar\omega)\Sigma_\omega^<(E) G_0^r(E)^\dag 
                        + G_\omega^r(E)\Sigma_0^<(E) G_0^r(E)^\dag.
\end{equation}
In the WBL, the DC lesser self-energy takes the form $\Sigma_0^<(E) = i \Gamma f_c(E)$, where $f_c(E)$ are the Fermi-Dirac distributions for the contacts. By Eq.~\ref{eq:sigma}, the AC self-energy is then given by $\Sigma_\omega^< = -\frac{ieV_{AC}\Gamma }{\hbar\omega} (f(E) - f(E+\hbar\omega))$. 

Observables can then be calculated from the lesser AC Green function in a fashion similar to DC NEGF. The frequency-dependent electron density $n_\omega(\v r)$ is given as
\begin{equation}
  n_\omega(\v r) = -i\int\frac{ dE}{2\pi}\, G^<_{\omega; \v r, \v r} (E) .
\end{equation}
While the electron density is important for charge dynamics, the AC current density must be calculated to compute the dynamic magnetic field within the inductor, and it is given by
\begin{equation}
        J_{\alpha,\omega}(\v r) = - \frac{e}{\hbar} \int \frac{dE}{2\pi}  \left[ H_{\v r + a_0\v{\hat \alpha} , \v r} G^<_{\omega; \v r, \v r + a_0\v{\hat \alpha}}(E)- G^<_{\omega; \v r + a_0\v{\hat \alpha}, \v r}(E) H_{\v r, \v r + a_0\v{\hat \alpha} }\right]  \quad (\alpha = x,y,z).
\end{equation}
Lastly, the AC contact current in the WBL is computed from as
\begin{equation}
        I(\omega) = \frac{e}{\hbar} \int \frac{dE}{2\pi} \left[ f(E) - f(E + \hbar\omega)\right] 
        \text{Tr}\left[ \Gamma_L G_0^r(E+\hbar\omega) \Gamma_R  G_0^r(E)^\dag\right] .
\end{equation}

The AC NEGF method can be computationally expensive since two matrix inversions are required to obtain  $G^r(E)$ and $G^r(E+\hbar\omega)$ at each step of the energy integration. Recursive methods that obviate the need of a full matrix inversion, however, can speed up computation significantly.\cite{Lake1997}

%%%%%%%%%%%%%%%%%%%%%%%%%%%%%%%%%%%%%%%%%%%%%%%%%%%%%%%%%%%%%%%%%%
%%%%%%%%%%%%%%%%%%%%%%%%%%%%%%%%%%%%%%%%%%%%%%%%%%%%%%%%%%%%%%%%%%
%%%%%%%%%%%%%%%%%%%%%%%%%%%%%%%%%%%%%%%%%%%%%%%%%%%%%%%%%%%%%%%%%%
% \subsection{Self-consistency with Electrodynamics}
\subsection*{Self-consistency with electrodynamics} 

For situations where the operating frequency is much lower than the inverse of an electron's transit time across a device, the quasistatic approximation of the electrostatic potential using the solution of Poisson's equation provides adequate accuracy.\cite{Larsson2007,Kienle2010} Above these frequencies, a full solution of Maxwell's equation must be obtained to incorporate dynamic electromagnetic coupling. For inductors, however, whose operation is dependent on magnetic coupling of the currents in the device, Poisson's equation is also inadequate as it fails to capture the magnetic response of the magnetic response of the current density. Therefore, to capture both the electric charge effects and the magnetic inductive effects of the device, we require the fully dynamic solution of Maxwell's equations. While typical electrodynamics simulations solve directly for the electric field, $\v E$, and magnetic field, $\v B$, quantum mechanics relies on the vector and scalar potentials.\cite{Chew2014} Therefore, we solve directly for the scalar potential $V$ and vector potential $\v A$ in the frequency domain using the Lorenz gauge, where $\nabla\cdot \v A = -\frac{i\omega}{c} V$, resulting in the following governing equations: 
\begin{eqnarray}
  \left(\nabla^2 + \frac{\omega^2}{c^2} \right) V = -\frac{\rho}{\varepsilon}, \\
  \left(\nabla^2 + \frac{\omega^2}{c^2} \right) \v A = -\mu \v J.
\end{eqnarray}
Here, $\omega$ is the frequency of interest, $c$ is the speed of light, $\varepsilon$ is the electric permittivity, and $\mu$ is the magnetic permeability. The FDFD formulation solves these equations using finite differences on a Yee cell\cite{Yee1966,Luebbers1990} using the charge and current densities from the AC NEGF simulation. The electromagnetics domain is larger than the NEGF domain to accommodate absorbing boundary conditions that allow for field radiation and hinder the development of cavity modes.\cite{Rickard2002,Shin2012} To reach self-consistency of this solution with the AC NEGF equation, the scalar and vector potentials are input back into the AC NEGF equations until the difference between the scalar potential on successive iterations is less than 1 $\mu$V.

\bibliography{Inductor}

\section*{Acknowledgements}

The authors acknowledge support from the NSF under CAREER Award ECCS-1351871. M.J.G. thanks D.S. Green for valuable discussions. T.M.P. thanks M.J. Park and Y. Kim for helpful discussions. 

\section*{Author contributions statement}

M.J.G initiated the project. T.M.P. performed the numerical simulations and analytical calculations. T.M.P. and M.J.G. analyzed the data and wrote the manuscript together.

\section*{Additional information}

\noindent\textbf{Supplementary information} accompanies this paper

\noindent\textbf{Competing financial interests:} The authors declare no competing financial interests

% \begin{figure}[ht]
% \centering
% \includegraphics[width=\linewidth]{stream}
% \caption{Legend (350 words max). Example legend text.}
% \label{fig:stream}
% \end{figure}

% \begin{table}[ht]
% \centering
% \begin{tabular}{|l|l|l|}
% \hline
% Condition & n & p \\
% \hline
% A & 5 & 0.1 \\
% \hline
% B & 10 & 0.01 \\
% \hline
% \end{tabular}
% \caption{\label{tab:example}Legend (350 words max). Example legend text.}
% \end{table}

% Figures and tables can be referenced in LaTeX using the ref command, e.g. Figure \ref{fig:stream} and Table \ref{tab:example}.

\end{document}